# Performance of low vision individuals when selecting a target with head-pointing in virtual reality

Camille Bordeau[1], Célia Passerel[1], Ambre Denis-Noël[2], Jean-Baptiste Melmi[1], Marianne Vaugoyeau[1], Carlos Aguilar[3], Iliana Huyet[4], Caroline Topart[4], François Devin[4], Frédéric Matonti[4], Pierre Kornprobst [5], Eric Castet[1]

**Authors' affiliations:**

[1] Aix Marseille Univ, CNRS, CRPN, Marseille, France

[2] Aix Marseille Univ, CNRS, LPL, Aix-en-Provence, France

[3] 2Clubdes3, Nice, France

[4] Centre Monticelli Paradis d'Ophtalmologie, Marseille, France

[5] Université Côte d'Azur, Inria, Sophia Antipolis, France

**Word count**: 9846

**Funding information**: This work was supported by grants from the Carnot Cognition Institute, the NeuroMarseille Institute and the ANR (20-CE19-0018).

**Commercial relationships disclosures**: Bordeau Camille, None ;Passerel Célia, None; Denis-Noël Ambre, None; Melmi Jean-Baptiste, None; Vaugoyeau Marianne, None; Aguilar Carlos, None; Huyet Iliana, None; Topart Caroline, None; Devin François, None; Matonti Frédéric, None; Kornprobst Pierre, None; Castet Eric, None

**Abstract**

**Purpose**: To investigate psychophysically the ability of low vision individuals with central visual field loss (CFL) to perform a visually-guided pointing task in a virtual reality environment.

**Methods**:

Patients with CFL (n=25, ages = 67-90 years) and normally-sighted controls (n=26, ages = 67-85 years) had to select a target (2° diameter dot) with a head-contingent cursor (6° diameter reticle). Target selection occurred when target was validly pointed at for 1.5 seconds. Pointing was valid when target was inside an invisible pointer activation zone (PAZ) centered on reticle. Task difficulty was decreased by increasing PAZ diameter from 0.5° to 8°. Performance was assessed by measuring the time needed to select the target. The task was also performed with an array of three simultaneously-displayed cursors.

**Results**:

Selection times decreased (from 14.1 and 8.4 seconds for patients and controls respectively) with increasing PAZ diameter and reached a similar asymptote for both groups (1.4 seconds). The rate of this decrease was smaller for patients so that PAZ diameter needed for their best performance was much larger than PAZ diameter needed for controls' best performance (average: 3.48° vs 1.32°). In the three-reticle condition, both groups tended to use the cursor closer to the target.

**Conclusions**:

Patients with CFL are able to point at a 2° target thanks to head-pointing. Their performance can get close to controls' best performance by increasing PAZ size.

**Translational relevance**:

This research suggests guidelines to improve the accessibility of visually-guided pointing tools for human-machine interfaces designed for low vision individuals.

## 1. Introduction

Central visual Field Loss (CFL) is one of the major causes of low vision, with Age-related Macular Degeneration (AMD) being the most frequent cause. Approximately 200 million of individuals were affected by AMD around the world in 2020, leading to moderate or severe low vision for more than 6 million of them[1]. This area of lost vision in the central visual field is often called a scotoma, which corresponds to the portion of the macula where photoreceptors are damaged. Vision in the central visual field, compared to peripheral vision, has the highest visual acuity, spatial resolution, contrast sensitivity, and the lowest spatial uncertainty and crowding[2–5]. Individuals with CFL are constrained to use peripheral vision as they must place an object of interest in a non-damaged peripheral area of the retina. The ability to efficiently use eccentric viewing is thus crucial to minimize impairments in visual functions[6,7] and this implies adaptation of oculomotor, visuomotor, and visual spatial attention processes[8,9]. With sufficient adaptation, some patients preferentially use one or several specific areas of the non-damaged peripheral retina to perform visual tasks such as fixating a target[10,11], single word reading[12] or pointing at targets[13]. These specific areas are called preferred retinal loci (PRL)[14].

The pragmatic implications of CFL in everyday life are numerous and serious especially when it comes to reading[6] and recognizing faces[15]. In short, everyday activities that rely on spatial resolution or contrast sensitivity are impaired: recognizing images of natural scenes[16], driving (Decarlo et al., 2003), using a computer[18–20], precise motor tasks such as making a sandwich[21], tracing a line[22], reaching and grasping an object[23–29], replacing an object at a specific location[27,28], or pointing at an object on a screen[13] – see [30] for a review on the impact of AMD on real-world visual abilities.

In this broad context of impaired visual abilities, our aim is to specifically investigate the ability of persons with low vision to point at objects. In the present work, "visually-guided pointing at an object" is defined as an action where body parts, essentially head, gaze or hand, are moved to achieve a visual alignment between a pointer and an object of interest[31,32]. In the commonly-studied case of hand-pointing, the pointer is usually the finger tip[33] or a contingent visual stimulus such as a mouse cursor which is another name for a hand-contingent stimulus[34,35]. The important idea here is that visually-guided pointing usually requires a high visual acuity because the object to be pointed at often has a small angular value. Think for instance of pointing at a small face picture with a handheld magnifier[36]. This is also clearly the case when individuals with low vision want to interact with tablets, computers or cell phones for instance to select items in menus[18–20,37]. In general, being able to point at targets seems a key requirement to be autonomous when using digital interfaces such as computers, smartphones, tablets or Virtual Reality (VR). Therefore, improving the accessibility of digital interfaces to low vision individuals[38,39], especially by improving pointing accuracy and speed[40], is a major challenge.

We address this challenge by investigating psychophysically the ability of participants with CFL to point at targets that need to be selected in VR environments (see below the advantages of using VR). This research axis will allow us to provide quantitative recommendations to render digital interfaces more accessible to persons with low vision. One important principle of our psychophysical approach is to study parametrically some of the successful techniques that have been used to improve pointing in normally-sighted persons having different degrees of motor impairments[41–51].

One of these famous and efficient pointing techniques was introduced in 1995 and is known as the "area cursor" technique[46]. In the initial work of Kabbash & Buxton, the "cursor" refers to a visible arrow that is contingent on mouse movements. In the standard pointing mode, selection of the target occurs when the arrow's tip is placed over the target. The novelty of the "area cursor" technique is that selection occurs when an area surrounding the arrow is placed over the target, thus increasing the effective size of the target in accordance with Fitt's law[45]. Our prediction is that this technique should be very efficient for persons with low vision, and we studied this efficiency as a function of the "area cursor's" size. In the present work, the "pointer activation zone or PAZ" expression is used, rather than "area cursor" expression, to emphasize a/ that this zone is not a visible contingent feedback but an invisible active zone allowing a pointing action and b/ that this zone pertains to any pointer (and not only to a hand-held mouse). In other words, the cursor (the visual stimulus providing a feedback contingent on the pointer's movement – note that this is a white 6° diameter reticle in our work, see Figure 1), should not be confused with the PAZ (the area or zone that is around the cursor and that allows to define whether a target is selected or not).

In addition to this famous "area cursor" technique, we also investigated for exploratory purposes if pointing could be correctly achieved, or even improved, by presenting three cursors, each having its PAZ, instead of a single cursor as used in all standard pointing tasks (see Figure 2). We wanted to test the hypothesis that the distance between the target and the cursor could be an important determinant of pointing performance. For instance, when a menu is displayed to the left, would it be helpful, to present the cursor to the left (as opposed to the right) thus providing a relatively small distance between target and cursor ?

As already briefly mentioned, the psychophysical investigation of pointing is achieved here in a VR environment, although the results of our research might be applied to a broader set of human-machine interfaces. VR was used in this study for three main reasons. First, VR offers an efficient and cheap way of simulating complex interactions such as pointing tasks within a 3D world[52]. These pointing tasks can be controlled with a great level of complexity and flexibility thus allowing us to study pointing performance with psychophysical methods (see our Methods section). Second, a VR experimental setup is also a very convenient clinical tool when implemented in a headset as it is easily transported and installed in hospitals or rehabilitation centers. Third, VR offers innovative ways of designing visual aids or visual rehabilitation protocols for persons with low vision. For instance, our research group has recently provided evidence that augmented vision techniques, similar to those initially developed as proofs of concept[53,54], can be incorporated in VR to help patients with low vision to efficiently select and augment some key stimuli such as faces[55].

We initially decided to investigate head-pointing performance (rather than eye- or hand-pointing) in the present work for different kinds of reasons. To our knowledge, head-pointing by persons with low vision has only rarely been investigated although it seems a promising approach[56]. In addition, measuring head movements has become relatively easy and cheap thanks to VR headsets, and this contrasts with measuring online eye movements of persons with low vision, a task which is notoriously difficult and requires expensive and sophisticated equipment[54,57]. Finally, concerning the choice between head- and hand-pointing, head-pointing was preferred after considering the following points. Humans naturally orient their head, and their spatially selective attention, toward objects of interest in their surrounding environment[58]. We felt that this should be an important aspect to consider especially because most elderly people with low vision in our pilot VR studies seemed to consider head-pointing as much more natural and easy than hand-pointing. Based on this preference, we therefore chose to first investigate head-controlled pointing in the present study while leaving the investigation of hand-pointing to a later study (which is currently ongoing in our lab).

To summarize, in the present experiments implemented in a VR headset, individuals with CFL and normally-sighted controls were recruited and tested in a visually-guided pointing task using a head-contingent pointer. Our goal was to produce a stringent psychophysical description of the pointing performance of individuals with CFL. This description should be useful a/ to provide a better theoretical understanding of the specific visuo-motor and attentional processes affected by CFL and b/ to issue recommendations allowing a better design of the pointing tools implemented in digital interfaces for persons with CFL or more generally with visual deficiencies[56].

## 2. Method

### 2.1. Participants

Twenty five patients with CFL (patients group, 10 women, age: Mean = 79.0; SD = 6.3, Range [67, 90]) and twenty six age-matched normally-sighted control individuals (control group, 20 women, age: Mean = 73.9; SD = 4.3, Range [67, 85]) participated in the study. Two patients were excluded from the study after their experimental session (one patient was not able to do the task, and the other one had a neurological disease). Individual characteristics of the included patients and control groups at the end of the inclusion process (patients: N = 23, control : N = 26) are summarized in Table 1 and Table 2, respectively. Patients with CFL were recruited at the Monticelli Ophthalmology Centre in Marseille, and were selected on the basis of diagnosis and visual acuity criteria. They were included in the study if (a) they had been diagnosed with a CFL pathology (22 wet AMD patients and 1 patient with a pachychoroid disease including pachydrusen), (b) if they had not been diagnosed with any other severe visual disease (eg, no glaucoma), and (c) if their eye with the worst visual acuity was between 0.5 and 1.4 logMAR (corresponding to letter x-height sizes between 0.26° and 2.09° - see [59]). Age-matched control participants were recruited at the Free time University of Aix-Marseille University. They were included in the study if they had not been diagnosed with a severe visual disease (eg, no CFL pathology nor glaucoma). Visual acuity was measured at the beginning of the experimental session by an orthoptist with the 4-m Monoyer scale.

All participants performed the task monocularly with their habitual refractive corrections. Patients with CFL performed the task with their worst acuity eye (a black screen was presented to the best eye within the VR headset), while control individuals performed the task with their best acuity eye (a black screen was presented to the worst eye within the VR headset). When both eyes of controls had the same visual acuity, the best eye was defined as the dominant eye.

The absence of severe or moderate cognitive impairment was checked with the Mini Mental State Evaluation (MMSE)[60] and the MMSE-blind version[61] for the control and patients groups respectively. The study was approved by the *Comité de protection des personnes Est II*. All participants gave their informed consent before being enrolled in the experiment, and the study followed the tenets of the Declaration of Helsinki.

*Table 1. Individual characteristics of the patients group. Visual acuity of the right eye (RE) and left eye (LE) is given with Snellen and logMAR notations. The asterisk symbol (*) indicates the eye used in the monocular viewing task (worse eye). F: female; M: male.*

| Patient (age, sex) | Visual acuity | | Diagnosis of CFL pathology | Number of valid trials (percentage of 120 total trials) |
|---|---|---|---|---|
| | RE (Snellen, logMAR) | LE (Snellen, logMAR) | | |
| SP01 (78, M) | 20/100 (0.7)* | 20/40 (0.3) | Wet AMD | 120 (100%) |
| SP02 (72, M) | 20/25 (0.1) | 20/80 (0.6)* | Wet AMD | 102 (85%) |
| SP03 (80, M) | 20/125 (0.8) | 20/500 (1.4)* | Wet AMD | 65 (54.2%) |
| SP06 (71, F) | 20/125 (0.8)* | 20/20 (0) | Wet AMD | 81 (67.5%) |
| SP07 (72, F) | 20/125 (0.8)* | 20/40 (0.3) | Wet AMD | 104 (86.7%) |
| SP09 (77, F) | 20/20 (0) | 20/400 (1.3)* | Wet AMD | 69 (757.5 %) |
| SP10 (80, F) | 20/200 (1)* | 20/32 (0,2) | Wet AMD | 98 (81.7%) |
| SP11 (90, F) | 20/160 (0.9) | 20/300 (1.2)* | Wet AMD | 63 (52.5%) |
| SP12 (90, M) | 20/100 (0.7) | 20/200 (1)* | Wet AMD | 82 (68.3%) |
| SP13 (67, F) | 20/100 (0.7)* | 20/50 (0.4) | Wet AMD | 113 (94.2%) |
| SP14 (85, M) | 20/400 (1.3)* | 20/50 (0.4) | Wet AMD | 97 (80.8%) |
| SP15 (81, M) | 20/50 (0.4) | 20/400 (1.3)* | Wet AMD | 97 (80.8%) |
| SP16 (77, M) | 20/50 (0.4) | 20/80 (0.6)* | Wet AMD | 1190 (99.2%) |
| SP17 (83, M) | 20/25 (0.1) | 20/100 (0.7)* | Wet AMD | 119 (99.2%) |
| SP18 (84, F) | 20/125 (0.8) | 20/400 (1.3)* | Wet AMD | 83 (69.2%) |
| SP19 (79, M) | 20/250 (1.1)* | 20/80 (0.6) | Wet AMD | 94 (78.3%) |
| SP20 (72, M) | 20/40 (0,3) | 20/100 (0,7)* | Wet AMD | 104 (86.7%) |
| SP21 (70, M) | 20/160 (0.9)* | 20/32 (0.2) | Pachychoroid / pachydrusens | 101 (84.2%) |
| SP22 (80, F) | 20/80 (0.6) | 20/500 (1.4)* | Wet AMD | 98 (81.7%) |
| SP23 (81, F) | 20/40 (0.3) | 20/63 (0.5)* | Wet AMD | 118 (98.3%) |
| SP24 (86, M) | 20/100 (0.7)* | 20/40 (0.3) | Wet AMD | 118 (98.3%) |
| SP25 (86, M) | 20/125 (0.8) | 20/200 (1)* | Wet AMD | 113 (94.2%) |
| SP26 (77, F) | 20/200 (1)* | 20/32 (0.2) | Wet AMD | 90 (75%) |

*Table 2. Individual characteristics of the control group. Visual acuity of the right eye (RE) and left eye (LE) is given with Snellen and logMAR notations. The asterisk symbol (*) indicates the eye used in the monocular viewing task (best eye). F: female; M: male.*

| Control (age, sex) | Visual acuity | | Auto-reported visual disease (eye) | Number of valid trials (percentage of 120 total trials) |
|---|---|---|---|---|
| | RE (Snellen, logMAR) | LE (Snellen, logMAR) | | |
| SCM01 (67, F) | 20/32 (0.2) | 20/32 (0.2)* | . | 118 (98.3%) |
| SCM02 (73, F) | 20/32 (0.2)* | 20/100 (0.7) | Cataract (LE) | 120 (100%) |
| SCM03 (72, F) | 20/25 (0.1)* | 20/50 (0.4) | . | 120 (100%) |
| SCM04 (72, F) | 20/20 (0)* | 20/25 (0.1) | . | 120 (100%) |
| SCM05 (71, H) | 20/40 (0.3)* | 20/63 (0.5) | . | 112 (93.3%) |
| SCM06 (72, H) | 20/32 (0.2)* | 20/40 (0.3) | . | 120 (100%) |
| SCM07 (81, H) | 20/20 (0)* | 20/20 (0) | . | 120 (100%) |
| SCM08 (72, F) | 20/40 (0.3) | 20/25 (0.1)* | Cataract (RE) | 120 (100%) |
| SCM09 (85, F) | 20/25 (0.1)* | 20/32 (0.2) | . | 120 (100%) |
| SCM10 (77, F) | 20/32 (0.2) | 20/25 (0.1)* | . | 119 (99.2%) |
| SCM11 (73, F) | 20/20 (0)* | 20/40 (0.3) | . | 119 (99.2%) |
| SCM12 (76, F) | 20/25 (0.1) | 20/20 (0)* | . | 108 (90%) |
| SCM13 (70, F) | 20/25 (0.1)* | 20/25 (0.1) | . | 120 (100%) |
| SCM14 (72, F) | 20/25 (0.1)* | 20/40 (0.3) | Cataract (LE) | 119 (99.2%) |
| SCM15 (70, F) | 20/40 (0.3)* | 20/50 (0.4) | . | 120 (100%) |
| SCM16 (69, F) | 20/25 (0.1)* | 20/32 (0.2) | . | 115 (95.8%) |
| SCM17 (68, F) | 20/25 (0.1)* | 20/25 (0.1) | . | 112 (93.3%) |
| SCM18 (76, F) | 20/63 (0.5) | 20/50 (0.4)* | . | 114 (95%) |
| SCM19 (74, H) | 20/25 (0.1)* | 20/25 (0.1) | . | 120 (100%) |
| SCM20 (72, F) | 20/25 (0.1)* | 20/32 (0.2) | . | 118 (99.2%) |
| SCM21 (77, H) | 20/32 (0.2) | 20/25 (0.1)* | . | 106 (88.3%) |
| SCM22 (78, F) | 20/25 (0.1)* | 20/40 (0.3) | . | 120 (100%) |
| SCM23 (74, F) | 20/100 (0.7) | 20/40 (0.3)* | . | 117 (97.5%) |
| SCM24 (71, F) | 20/20 (0)* | 20/20 (0) | . | 120 (100%) |
| SCM25 (79, H) | 20/32 (0.2) | 20/32 (0.2)* | Cataract (RE) | 119 (99.2%) |
| SCM26 (80, F) | 20/25 (0.1) | 20/25 (0.1)* | Cataract (LE) | 108 (90%) |

## 2.2. Experimental setup

Experiments for patients and controls took place respectively at the Monticelli Paradis Opthalmogy Center and at Aix Marseille University. The VR experiment was developed with the Perception Toolbox for Virtual Reality (PTVR)[52] ; details about PTVR can also be found in its documentation (https://ptvr.inria.fr/). The python codes to run the experiment will be available in open access in PTVR release 2.1.2 upon acceptance (Assets > Other > “Experiment codes” - https://gitlab.inria.fr/PTVR_Public/PTVR_Researchers/-/releases/PTVR_2.1.2). The availability of our research program code should help other teams reproduce and extend the present work[62].

Participants sat in a comfortable chair with adjustable height, and were wearing an HTC Vive Pro Eye head-mounted display (HMD) and holding a VR handcontroller. The nominal size of the field of view of the HMD was 110°. Headset and controller were tracked with two base stations located on both sides of the participant’s chair. The best, or worse, eye of patients with CFL, and respectively controls, was prevented from viewing by blanking out the corresponding monocular screen in the VR headset.

## 2.3. Stimuli and head-pointing selection task

Positions of stimuli were specified in two different azimuth-elevation coordinate systems (the acronym « CS » will be sometimes used instead of « Coordinate System » in the present work for the sake of clarity). One coordinate system was static with respect to the real world (world-centered CS). The second coordinate system was head-contingent (head-centered CS). Stimuli that were static in the virtual environment (e.g. the target) were specified in world-centered coordinates. Stimuli that were head-contingent (the reticles) were specified in head-centered coordinates. The origin of the world-centered CS was 1.1 m above the floor at the location of the chair on which the participant was sitting on, thus close to the participant’s eyes (this position is referred to as the « viewpoint » when the participants’ eyes are located at this position).

At the beginning of each trial, a static target and two static vertical rectangles appeared simultaneously with one or three head-contingent cursors. The static target was displayed on one of the two vertical rectangles. This is representend in Figure 2A in a participant's-eye-level view and in Figure 2B in a bird's-eye view. The two grey rectangles (6.0 cd/m²; width = 10°; height = 30°) were on the left side (rectangle's world-centered coordinates : azimuth = -20°; elevation = 0°) and on the right side (rectangle's world-centered coordinates: azimuth = +20°; elevation = 0°). Each vertical rectangle was perpendicular to the line joining the rectangle's center to the origin of the world-centered CS. The distance between these stimuli and the origin of the world-centered CS (often referred to as the « viewing distance ») was 495 meters.

The target was a 2°-diameter black disk horizontally-centered on its underlying rectangle and its vertical position was at one of five elevations ( 0°, ±5° or ± 10°). In Figure 2A, the target's world-centered coordinates are azimuth = -20°, elevation = +5°. Black stimuli luminance was 0.1 cd/m².

When the participants' eyes are located at the origin of the world-centered CS, it is convenient to express the position of stimuli in a perimetric CS (with the same CS origin). In this case, the left rectangle's world-centered coordinates were an eccentricity of 20° and a half-meridian = 180°, while the right rectangle's wolrd-centered coordinates were an eccentricity of 20° and a half-meridian of 0° (viewing distance was still 495 meters). The background color of the virtual environment was grey (3.3 cd/m²). The cursor was a head-contingent white reticle that had a 6°-diameter with 4 inside bars of 1° of length (Figure 1). White stimuli luminance was 114 cd/m². As mentioned in the introduction, the cursor is the visual stimulus providing a feedback contingent on the pointer's movement (here head movement). The number of displayed reticles was either one or three. Their coordinates were specified in a head-centered coordinate system. When only one reticle was displayed, its coordinates were: azimuth = 0°, elevation = 0°. When three reticles were displayed, two additional reticles were added with their head-centered azimuth at ± 10° and elevation at 0° (Figure 2A). In the perimetric coordinate system, the left reticle's head-centered coordinates were: eccentricity = 10° and half-meridian = 180°, while the right reticle's head-centered coordinates were: eccentricity = 10° and half-meridian = 0°. Each reticle was perpendicular to a line joining its center to the origin of the head-centered coordinate system (Figure 2B).

Participants performed a point-and-dwell selection task. The participants' task was to move their head, and thus the reticle(s), to point at the target disk by aligning as much as possible the center of the reticle (or one of the reticles) with the center of the target disk. A video example illustrating the task is available in the PTVR Youtube channel (https://www.youtube.com/watch?v=FUCIsH337vU&list=PL2eK3zoW1LsK8CVWRNP72iarfF_dW48VN&index=3) - note that spatio-temporal parameters of stimuli used in the experiment are different from those depicted in the video. This is represented in Figure 2C where the center of the middle reticle is perfectly aligned with the center of the target. A correct alignment, or similarly a correct pointing, was defined thanks to a circular tolerance zone centered on the reticle (referred to hereafter as the Pointer Activation Zone – PAZ) – See Figure 15 in Castet et al. (2024)[52]. As mentioned in the introduction, the PAZ corresponds to the "area cursor" introduced by Kabbash & Buxton (1995)[46]. The PAZ is the area centered on the reticle that allows to define whether the target is selected or not, and should not be confused with the reticle itself. In our study, PAZ diameter was modulated to vary the effective size of the target, and thus the difficulty of the task[46].

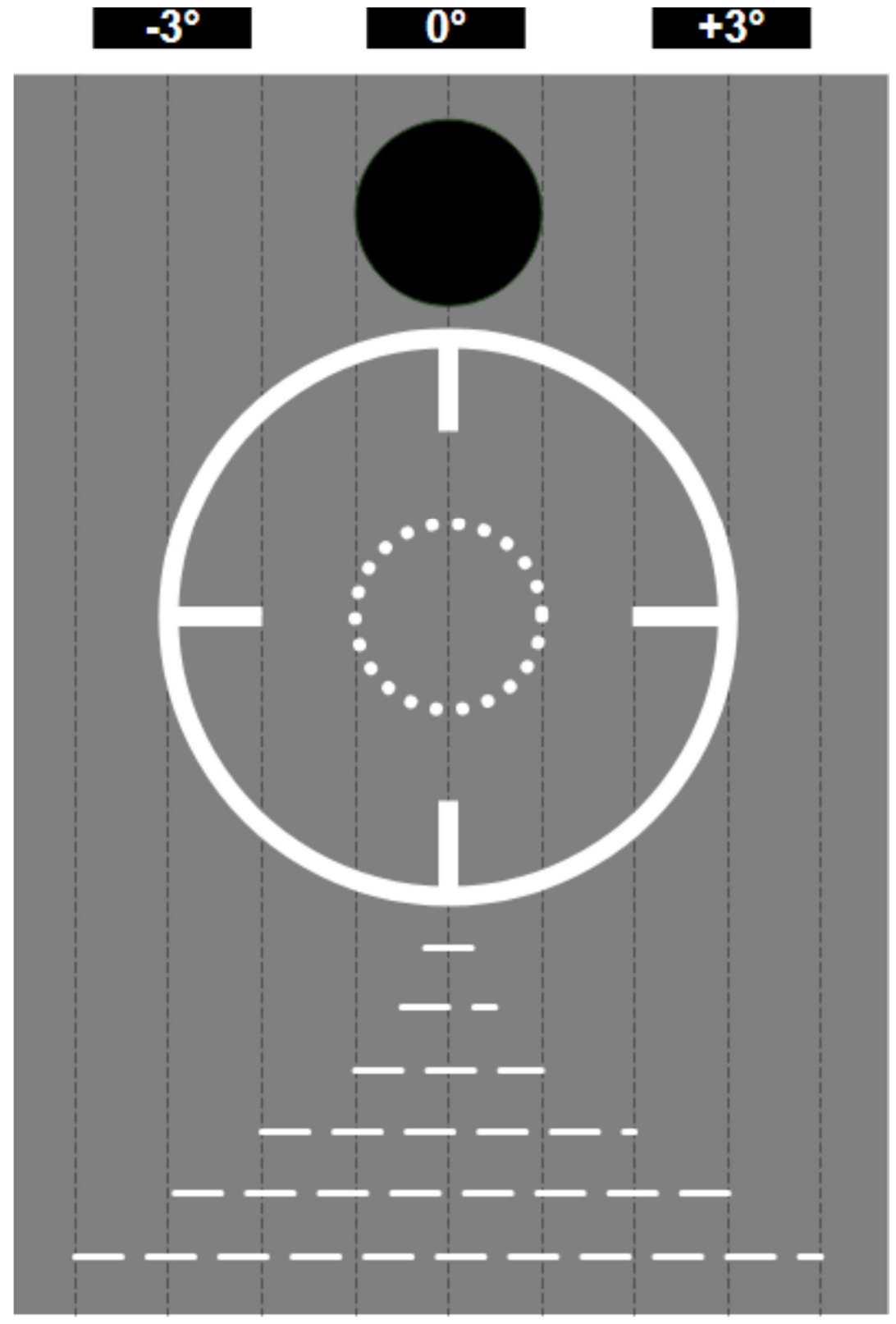

*Figure 1. Sizes of target (black disk), cursor (white reticle) and pointer activation zone (PAZ) (white dotted circle within the reticle). Diameters of target and of reticle are respectively 2 and 6 degrees. The six PAZ diameters that were tested are represented by white horizontal dashed lines (0.5, 1, 2, 4, 6, and 8 degrees). Dashed and dotted lines were not displayed in actual experiments.*

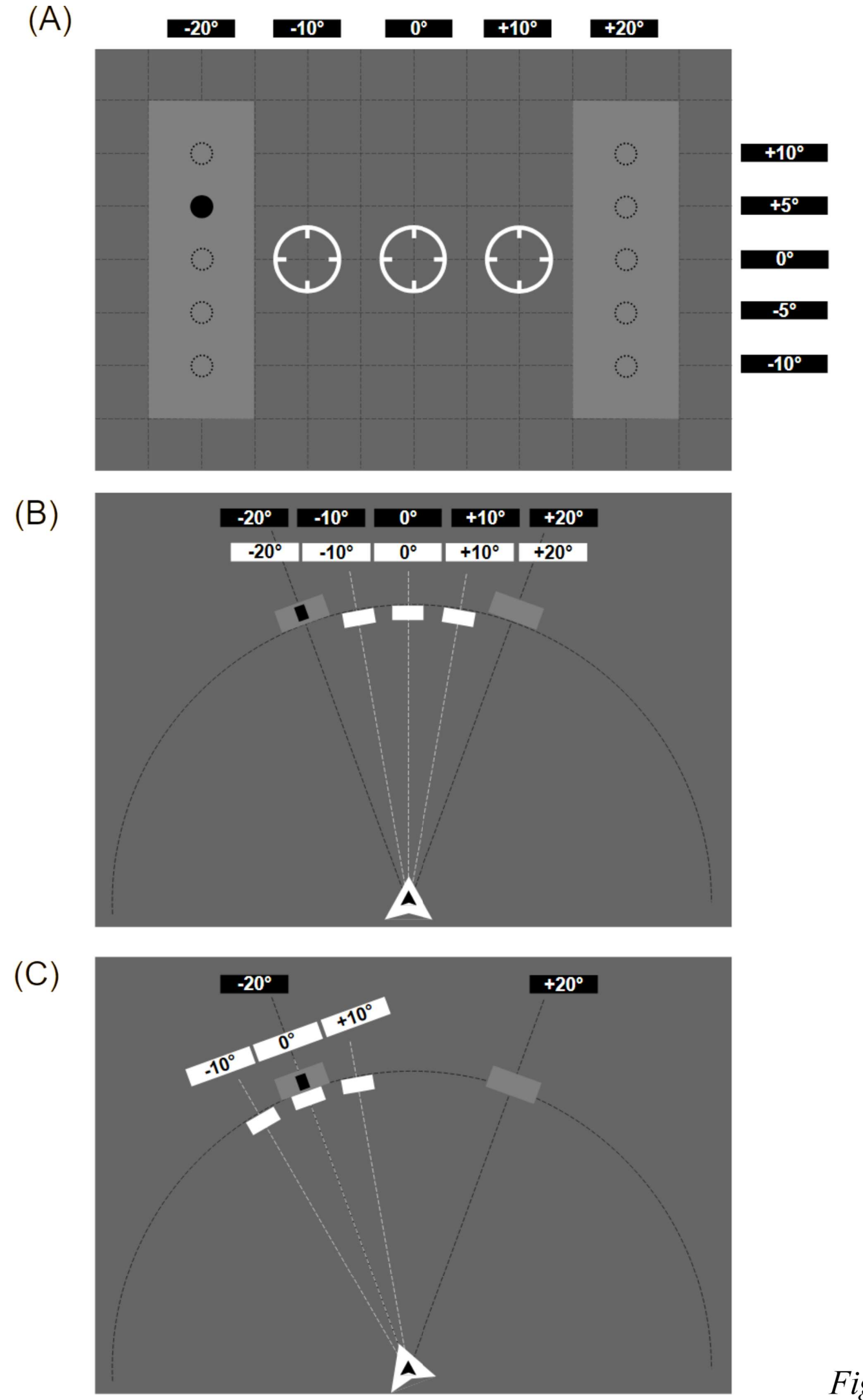

*Figure 2. Positions of the stimuli. Head-centered and world-centered coordinates are respectively displayed within white and black rectangular labels (azimuth-elevation coordinate system). Black dashed lines were not displayed in actual experiments. (A) Participant's-eye-level view: possible positions of the target (black disk) on one of the two grey vertical « rectangles » are represented by black dashed circles. The three head-contingent cursors (white reticles) used in the 3-reticle condition are also depicted. (B) Bird's-eye view with head straight-ahead. (C) bird's-eye view with head to the left (azimuth: -20°). Here the head is perfectly pointing at the target. In (B) and (C), the origins of the world-centered and head-centered coordinate systems are respectively represented by the black and white triangles (not displayed in actual experiments). The target, reticles, and rectangles are respectively represented by black, white and grey bars.The grey and white bars were displayed at the same viewing distance, although they are slighly separated in (B) and (C) for visual clarity.*

An example where the PAZ diameter is 2° is shown in Figure 1. The six possible PAZ diameter values were 0.5°, 1°, 2°, 4°, 6° and 8°. The PAZ was not visible in actual experiments, and participants were not explicitly aware of its existence. Pointing was considered as correct when the PAZ contained the target center. For instance, Figure 3A shows an example where the PAZ does not contain the target's center, so that pointing is not considered as correct, whereas Figure 3B shows an example where pointing is considered as correct since the PAZ contains the target's center. Figure 3C illustrates an experimental condition where pointing is very easy as the PAZ is very large (8°) whereas in Figure 3D pointing is very difficult as the PAZ is very small (0.5°).

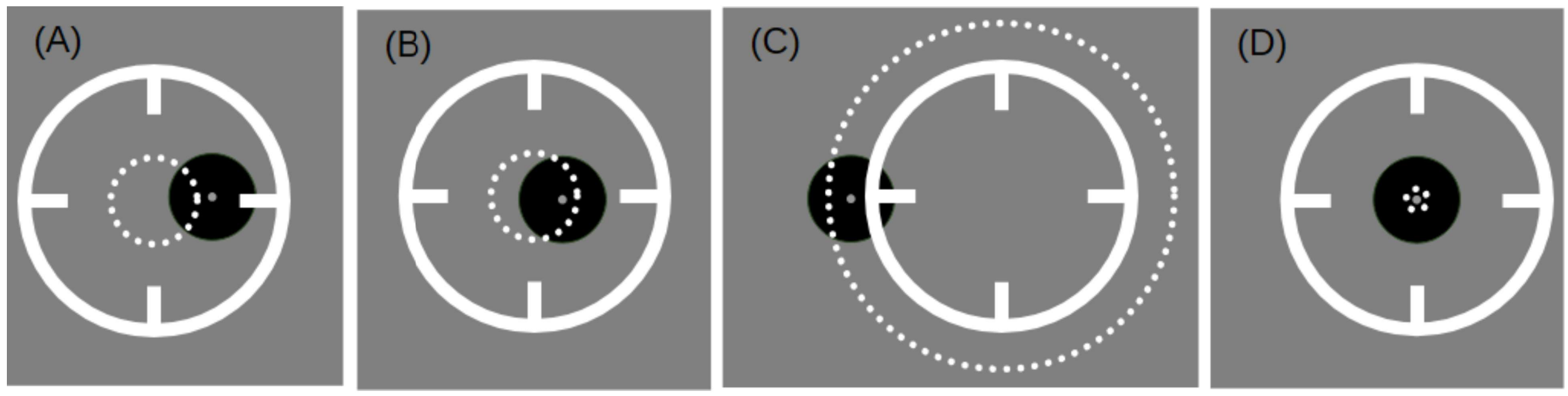

*Figure 3. How does the Pointer Activation Zone (PAZ) determine if pointing is correct or not. The center of the target (black disk) is represented by a tiny grey dot. Three examples of PAZ diameters are depicted with white dotted lines: 2° in (A) and (B), 8° in (C) and 0.5° in (D). The PAZ and the target center were not visible in actual experiments.(A) Pointing is not considered as correct as the center of the target is not within the PAZ.(B), (C) and (D) Pointing is considered as correct as the center of the target lies within the PAZ. As soon as pointing is correct, the target is said to be « pre-selected » (see text) and it starts flickering.*

Participants were informed when pointing was correct thanks to a visual feedback: as soon as the PAZ contained the target's center, the target started flickering (4 Hz-rate black and white alternation). When this happened, the target was said to be « pre-selected » in the sense that participants now needed to validate this correct selection by stabilizing their head (thus maintaining the visual flickering feedback) for a non-interrupted dwell-time of 1500 ms. As soon as this non-interrupted period of visual flickering had elapsed, the selection of the target was automatically validated and the trial was stopped and considered valid. Selection time was defined as the interval between target's appearance and trial end minus 1500 ms (the latter being the dwell-time, ie the duration of visual flickering that patients had to maintain without any interruption). Participants had to validate the selection within 30 s, otherwise the trial was stopped and was considered as a timeout trial. At the end of a trial, participants were warned with an audio message if a valid pointing or a timeout had occurred. When three reticles were displayed, participants were instructed to point at the target with the reticle they preferred and they were free to use any reticle across trials. In this condition, the reticle used to select the target was recorded in each trial.

### 2.4. Procedure

Participants were tested in a unique session lasting approximately 1h. The timeline of the session is given in Figure 4 and detailed below. After participants had given their informed consent, the experimenter explained the task and introduced the participants to the VR equipment. Once the participants were equipped with the HMD and the handcontroller, the test procedure started.

The test procedure consisted in a total of 120 trials per participant, distributed accros two successive phases (Phase 1 and Phase 2). Phases differed in the number of reticles : 1-reticle condition or 3-reticle condition. In the 3-reticle condition, the three reticles were simultaneously displayed (see Figure 2A), whereas in the 1-reticle condition, only the middle reticle was displayed. The order of the two phases was counterbalanced across the participants. Each phase started by a familiarization block (FB) of 10 trials during which participants were familiarized with the virtual environment and with the task. The familiarization block was followed by 6 experimental blocks (from B1 to B6) of 10 trials each.

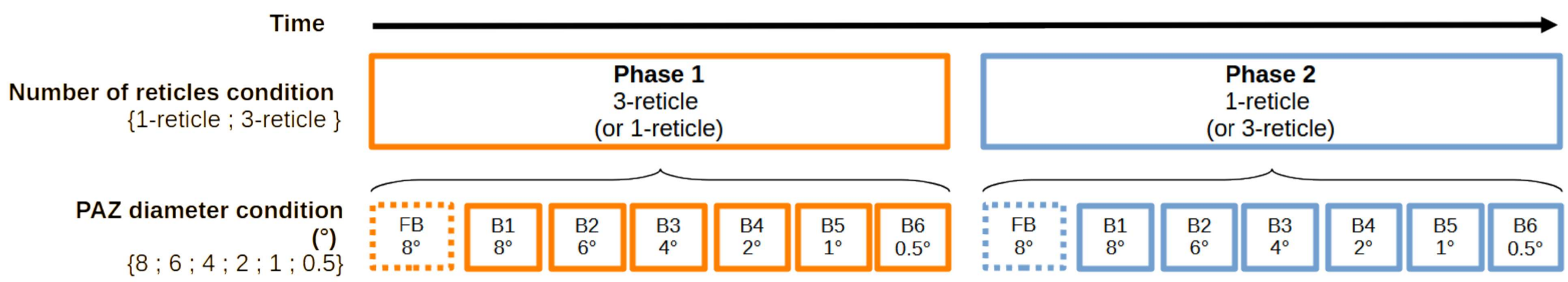

*Figure 4. Timeline of an experimental session. The 1-reticle and 3-reticle conditions were tested in two experimental phases (order of phases was counter-balanced across participants). Each Phase was divided into 7 blocks of 10 trials: one familiarisation block (FB, dotted lines) and six experimental blocks (B1 – B6). One PAZ diameter condition was tested in each experimental block in a decreasing order.*

In the FB blocks, the PAZ diameter was always 8°. Within each Phase, the six PAZ diameter values were tested in separate experimental blocks. The order of blocks was the same for all participants : decreasing order from 8° to 0.5° (corresponding to an increase of task difficulty). This experimental design is extensively used in clinical research with visually impaired patients, for instance when measuring visual acuity[63] or the critical print size in the MNREAD test[64].

An experimental block was stopped before the end of the 10 trials if 6 successive timeouts occurred within the block or if participants expressed the wish to stop the block. Ending a block before its end resulted in the end of the current Phase (lower values of PAZ diameter were thus not tested). The number of completed trials per participant (timeout excluded) in the patients and control groups are respectively mentioned in Table 1 and Table 2.

Before launching each trial, participants were instructed to orient their head as much as possible in the straight-ahead direction, and the experimental program checked that participants' head was oriented approximately straight ahead (an angular deviation of 4 degrees from straight-ahead was tolerated). The goal of this online control was to make sure that the average angle between head orientation at the beginning of a trial and the azimuth of the target was constant for targets on the left and on the right side. To help participants orient their head as required, the following procedure was used: a large black static straight-ahead cross (world-centered coordinates: azimuth = 0°; elevation = 0° ) and a large white head-contingent cross (head-centered coordinates: azimuth = 0°; elevation =

0°) were simultaneously displayed. The bars of the black cross had a width of 3° and a length of 8°, and the bars of the white cross had a width of 1.2° and a length of 8°. This is representend in Figure 5A and Figure 5B with respectively a participant's-eye-level view and a bird's-eye view.

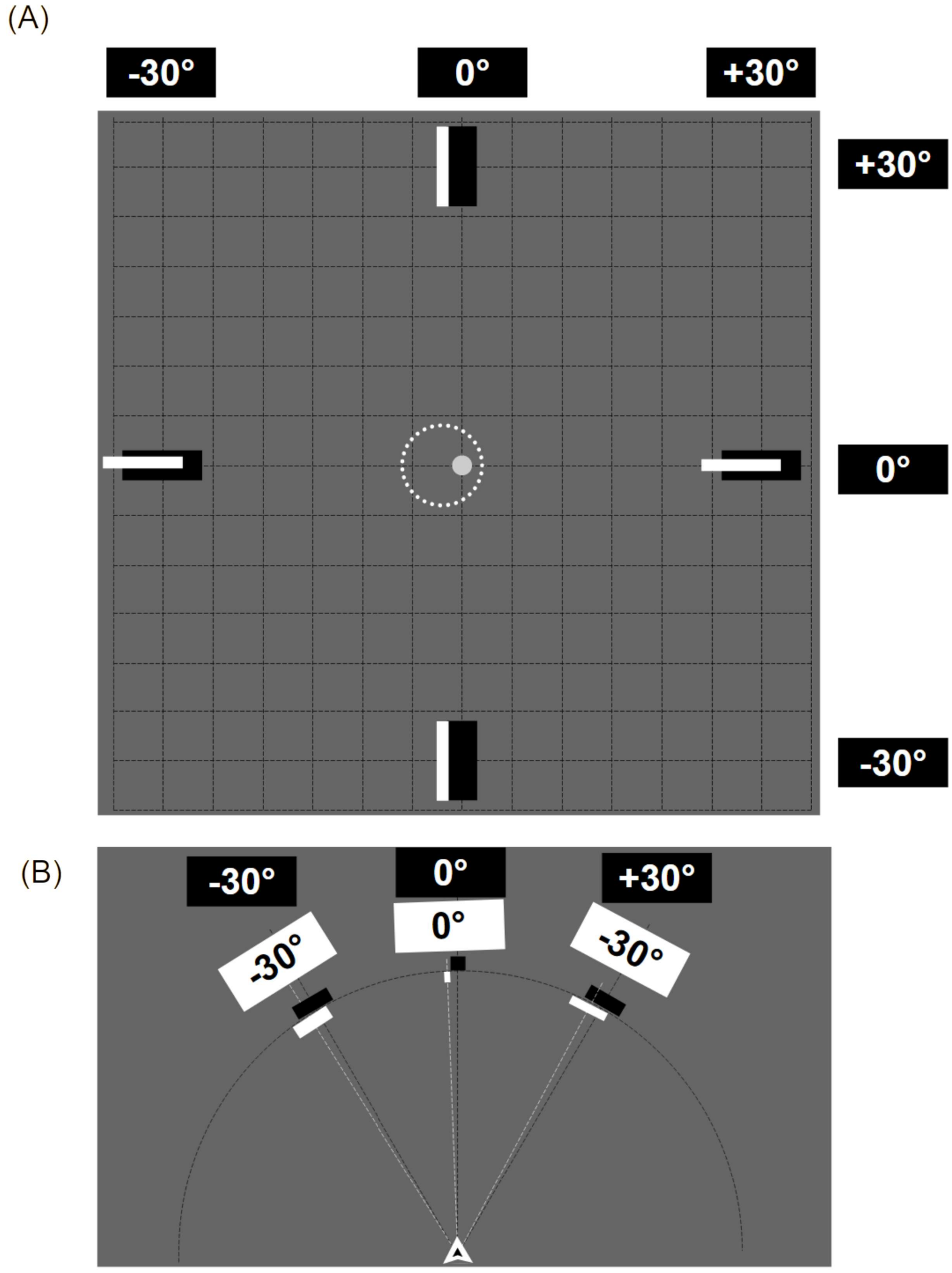

*Figure 5. Schematic illustration of the online control of participants' head orientation before each trial. Positions of the straight-ahead and head-contingent crosses are shown in a participant's-eye-level (A) and bird's-eye views (B). See Figure 2 for legend. In this example, the background of the screen is grey because the center (represented by the tiny grey dot) of the static straight-ahead cross (4 black bars) is within the circular tolerance zone (white dotted circle) of the head-contingent cross (4 white bars). Otherwise, if head orientation was not valid (see text), the background would be red.*

Participants were told that superimposing the white and the black crosses would help them orient their head straight-ahead. In addition, a negative visual feedback was provided to inform participants when their head orientation was not sufficiently straight-ahead : the background became red (5.5 cd/m²) when the head orientation deviated by more than 4 degrees from straight-ahead. In the latter case, the next trial could not be launched if participants pressed the handcontroller's trigger.

### 2.5. Statistical analyses

Statistical analyses were conducted with R[65] version 4.3.1.

#### 2.5.1. Selection times analysis

Selection times were analyzed with non-linear mixed-effects models (NLMM) using nlme package[66] version 3.1-162. Selection time was modeled as a function of the PAZ diameter following an exponential decay function using the program SSAsym():

$$ST' = SSAsym(PAZ) = Asym + (R0 - Asym) \times \exp^{-\exp^{lrc} \times PAZ}$$

with ST' the predicted selection time, PAZ the PAZ diameter, Asym the asymptote, R0 the predicted selection time value when PAZ is 0, and lrc the logarithm rate constant which reflects the rate of the decrease towards the asymptote (ie, the speed of convergence of the curve to the asymptote). PAZ diameter values were transformed by a translation of the origin (PAZ = PAZ – 0.5°). However, for clarity, the results of the analyses are presented with the original values of PAZ diameter in the Results section. Only trials in which no timeout occurred were included in the model, thus resulting in a total of 5293 trials (this produced between 63 and 120 trials per participant - see column « Number of trials completed » in Table 1 and Table 2). Model selection was performed using the Akaike's Information Criterion (AIC) and likelihood ratio test with the anova program of the nlme package[66]. Group (Patients vs Control) was specified as fixed factor while participants were specified as random factors. During the model selection process, including the number of reticles (1 vs 3) as fixed effect did not improve the predictability of the model. Thus this independent variable was not included in the final model. In addition, the effect of this variable on selection time was not significant, whatever the parameter (neither Asym, R0, nor lrc).

**2.5.2. Prefered reticle analysis in the 3-reticle condition**

In each trial in the 3-reticle condition, participants were free to use any of the three reticles to select the target. The prefered reticle was analyzed with a generalized linear mixed-effects model (GLMM). Only trials in which no timeout occurred were included in the model, thus resulting in a total of 2654 trials (this produced between 32 and 60 trials per participant). Depending on the azimuth of the target (world-centered azimuth = -20° for left side, and +20° for right side), the lateral reticles (head-centered azimuth = ± 10°) were categorized as ipsilateral or contralateral to the target. The reticle in the middle of the viewport (head-centered azimuth = 0°) was always categorized as the middle reticle. For instance, in Figure 2B, the reticle on the left side (head-centered azimuth = -10°) is categorized as the ipsilateral reticle because it is on the same side as the target, while the reticle on the right side (head-centered azimuth = +10°) is categorized as the contralateral reticle. The percentage of preference of the ipsilateral, contralateral, and middle reticles was modeled with the glmer function of the lme4 package[67] version 1.1.34, using binomial distributional family, and logit link function. Pairwise comparisons were conducted using the emmeans package[68] version 1.8.8. Model selection was performed using the Akaike's Information Criterion (AIC) and likelihood ratio test with the anova program of the nlme package[66]. Position (ipsilateral vs contralateral vs middle) and Group (Patients vs Control) were specified as fixed factors while participants were specified as random factors. During the model selection process, we checked that the azimuth position of the target (-20° vs +20°) has no significant effect on the prefered reticle. Including this variable as fixed effect did not improve the predictability of the model. Thus the variable “azimuth of the target” was not included in the final model. In addition, the effect of this variable on the prefered reticle was not significant. The GLMM model gives estimates in log-odds. However, for clarity, the results of the analyses are presented with percentage values in the Results section.

**3. Results**

### 3.1. Selection times

Average selection times of patients and controls are respectively 3682 ms ± 4864 (standard deviation) and 2692 ms ± 3294 (standard deviation). Minimal selection times of patients and controls are respectively 555 ms and 522 ms. The difficulty of the task was manipulated by varying the PAZ diameter in separate experimental blocks. Selection times as a function of PAZ diameter and number of reticles are presented separately for each participants of the patients and control groups respectively in Figure 6 and Figure 7. These figures suggest that selection times tend to decrease as PAZ diameter increases, up to an asymptote. This is confirmed by the fixed effects of the NLMM which are presented in Table 3 and plotted in Figure 8.

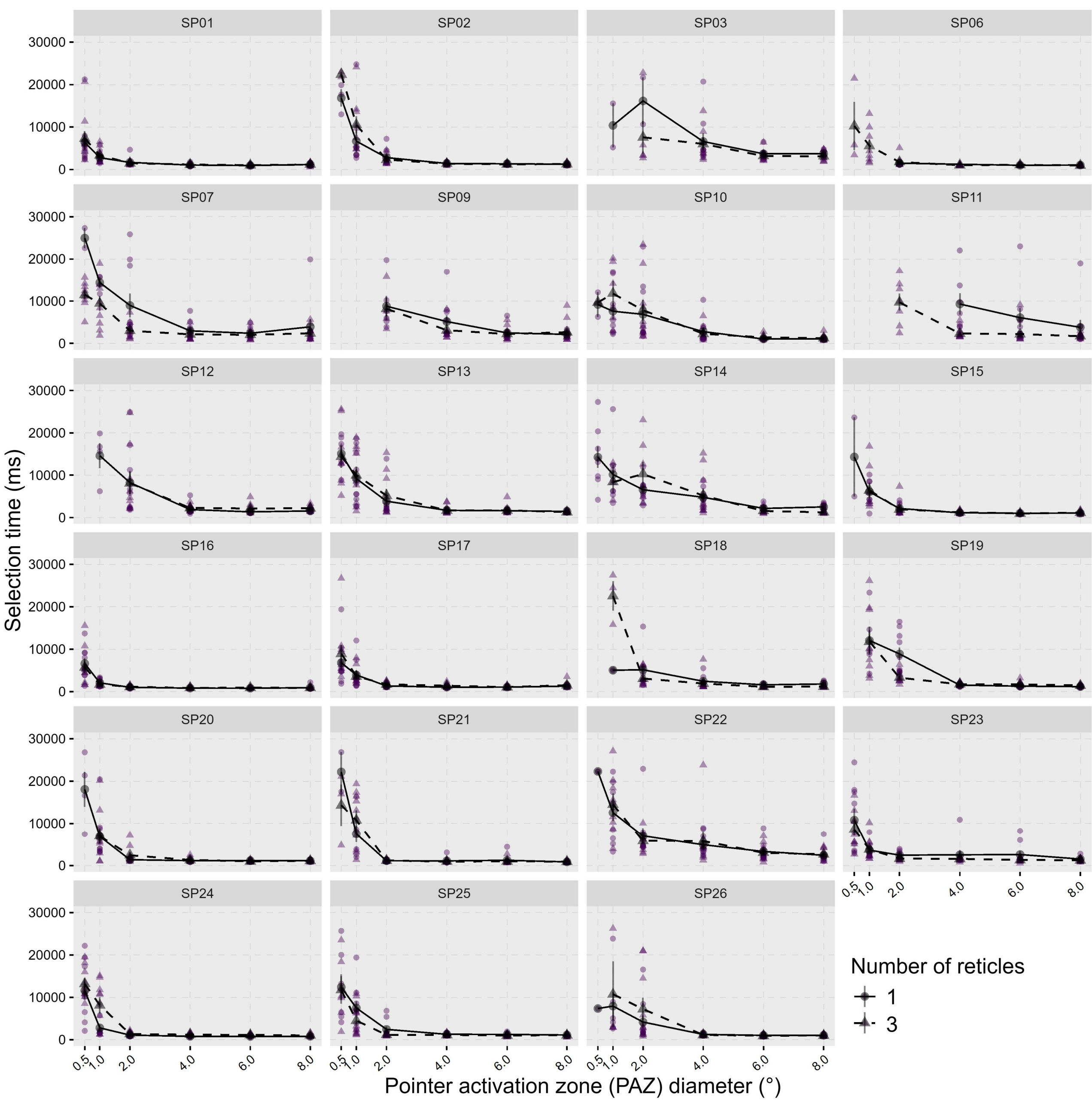

*Figure 6.* *Selection times as a function of PAZ diameter for each patient. Each facet represents one participant. Selection times measured in each trial are plotted for the 1-reticle condition (purple dots) and the 3-reticle condition (purple triangles). Average selection times are plotted separately for the 1-reticle condition (black dots, solid lines) and the 3-reticle condition (black triangles, dashed lines). Error bars represent standard errors.*

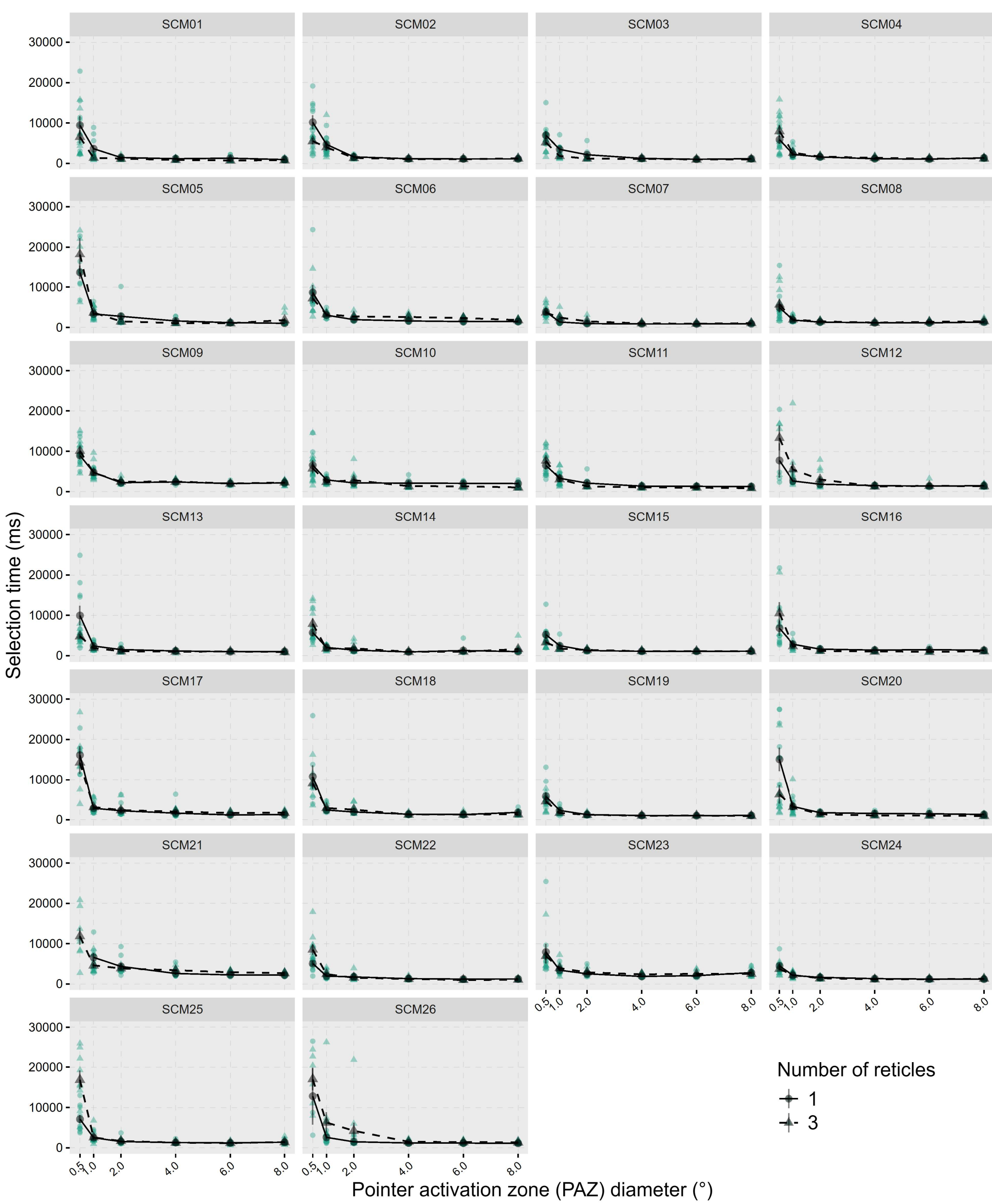

*Figure 7. Selection times as a function of PAZ diameter for each control participant. Each facet represents one participant. Selection times measured in each trial are plotted for the 1-reticle condition (green dots) and the 3-reticle condition (green triangles). Average selection times are plotted separately for the 1-reticle condition (black dots, solid lines) and the 3-reticle condition (black triangles, dashed lines). Error bars represent standard errors.*

*Table 3. Fixed effects estimates from the non-linear mixed-effects model. Asym, R0 and lrc respectively represent the asymptote, the selection time when PAZ is at its reference level (0.5), and the logarithm rate constant. The reference level of the « group » independant variable is the control group. SE : standard error. ***: p-value < .0001.*

| | Estimate | SE | degree of freedom | t-value | p-value |
|---|---|---|---|---|---|
| Asym Intercept | 1470.3 | 88.71 | 5239 | 16.574 | *** |
| Asym GroupPatient | -79.0 | 141.47 | 5239 | -0.5587 | .5764 |
| R0 Intercept | 8403.6 | 745.33 | 5239 | 11.2751 | *** |
| R0 GroupPatient | 5730.3 | 1125.80 | 5239 | 5.08996 | *** |
| lrc Intercept | 1.11 | 0.11 | 5239 | 9.6267 | *** |
| lrc GroupPatient | -1.1 | 0.16 | 5239 | -6.6962 | *** |

The asymptote (Asym) estimates of controls and patients are respectively 1470.3 ms and 1391.3 ms but the difference is not significant ($p = .5764$). The asymptote represents the best performance (smallest selection times) when PAZ increases. The R0 estimate for the control group is 8403.6 ms, which is significantly lower than the R0 estimate in the patient group (14133.9 ms, $p < .0001$). When taking into account the change of origin of the PAZ variable, the R0 estimate represents the predicted selection time for the smallest tested PAZ diameter (PAZ = 0.5°), namely when task difficulty was the highest. The effect of group on lrc (logarithm rate constant) was significant ($p < .0001$). The lrc estimate of the control group is 1.11 which is significantly lower than the lrc estimate of the patient group (lrc = 0.01). The lrc estimate represents the rate of the decrease towards the asymptote. The higher the lrc, the faster the decrease of selection times as PAZ diameter increases. In our case, being close to the asymptote reflects better performance (smaller selection times). This can be seen in Figure 8 where the rate of the decrease of the patients group is smaller than in the control group (population-level curves).

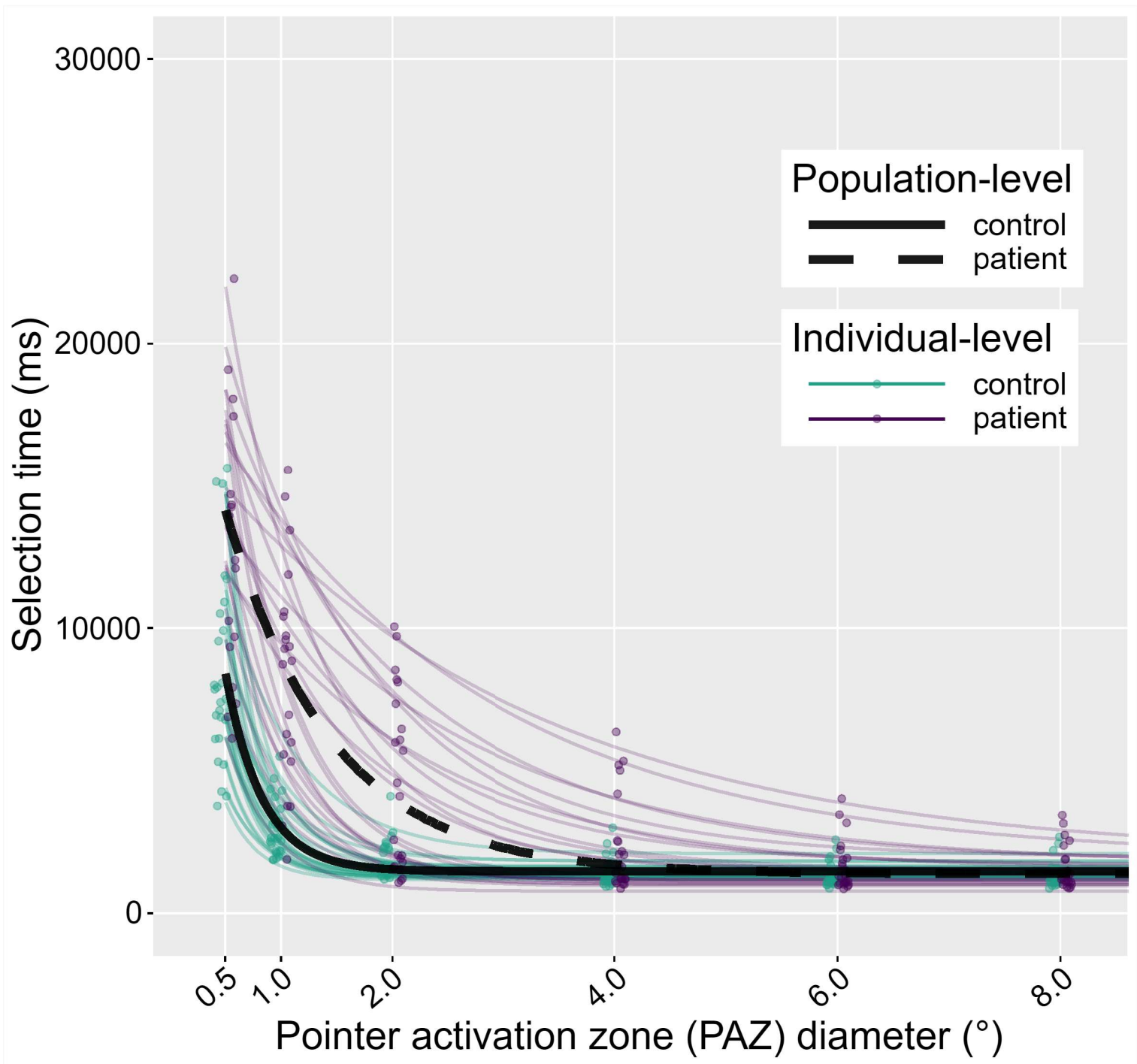

*Figure 8. Effect of PAZ diameter and group on selection time from the NLMM. Black lines represent the effect of PAZ diameter at the population level. Thin solid lines represent the effect of PAZ diameter for each participant of the patients group (dark purple lines) and of the control group (light green lines). Dots represent the across-trial average selection time measured for each participant (1- and 3-reticle conditions aggregated).*

To sum up, at the population level, selection times decreased as the task difficulty decreased (as PAZ diameter increased) for both groups. When task difficulty was maximal (smallest PAZ diameter of 0.5°), patients needed more time than controls to select the target (higher R0).-The rate of the decrease of selection times towards the asymptote was smaller in the patients group than in the control group (lower lrc). However, the minimum selection times (asymptotes) of patients and control groups were not significantly different.

We defined the Critical PAZ (CPAZ) diameter, as the PAZ diameter threshold value above which selection time starts remaining below a certain selection time criterion (STC). The CPAZ value is also useful to characterize the non-linear relationship between selection time and PAZ diameter with a unique index. The selection time criterion was defined as STC = Asym × Ɛ ms, with Asym the asymptote estimate, and Ɛ = 1.4. The CPAZ diameter was thus defined as:

$$CPAZ = \frac{-ln\left(\frac{(STC) - Asym}{(R0 - Asym)}\right)}{exp^{lrc}}$$

The STC and CPAZ diameter values at the individual level were extracted with the Best Linear Unbiased Predictions (BLUPs, see Pinheiro & Bates, 2000). The CPAZ diameter values of the patients and control groups are shown in Figure 9A**.** The STC of the patients and control groups were 1948 ms and 2058 ms, and their CPAZ diameter were 3.48° and 1.32°. As shown in Figure 9B, CPAZ diameter values varied across participants, especially in the patients group. CPAZ diameter values of the patients group varied between 1.7° and 8.7°, while values between 1° and 2° were observed in the control group. As an illustration, the patient SP16 has a CPAZ diameter of 1.7° while SP07 has a higher CPAZ diameter of 4.6°. The difference in CPAZ values between these two participants is clearly visible in Figure 6: the rate of decrease of selection times towards the asymptote in the panel of SP07 is smaller than in the panel of SP16.

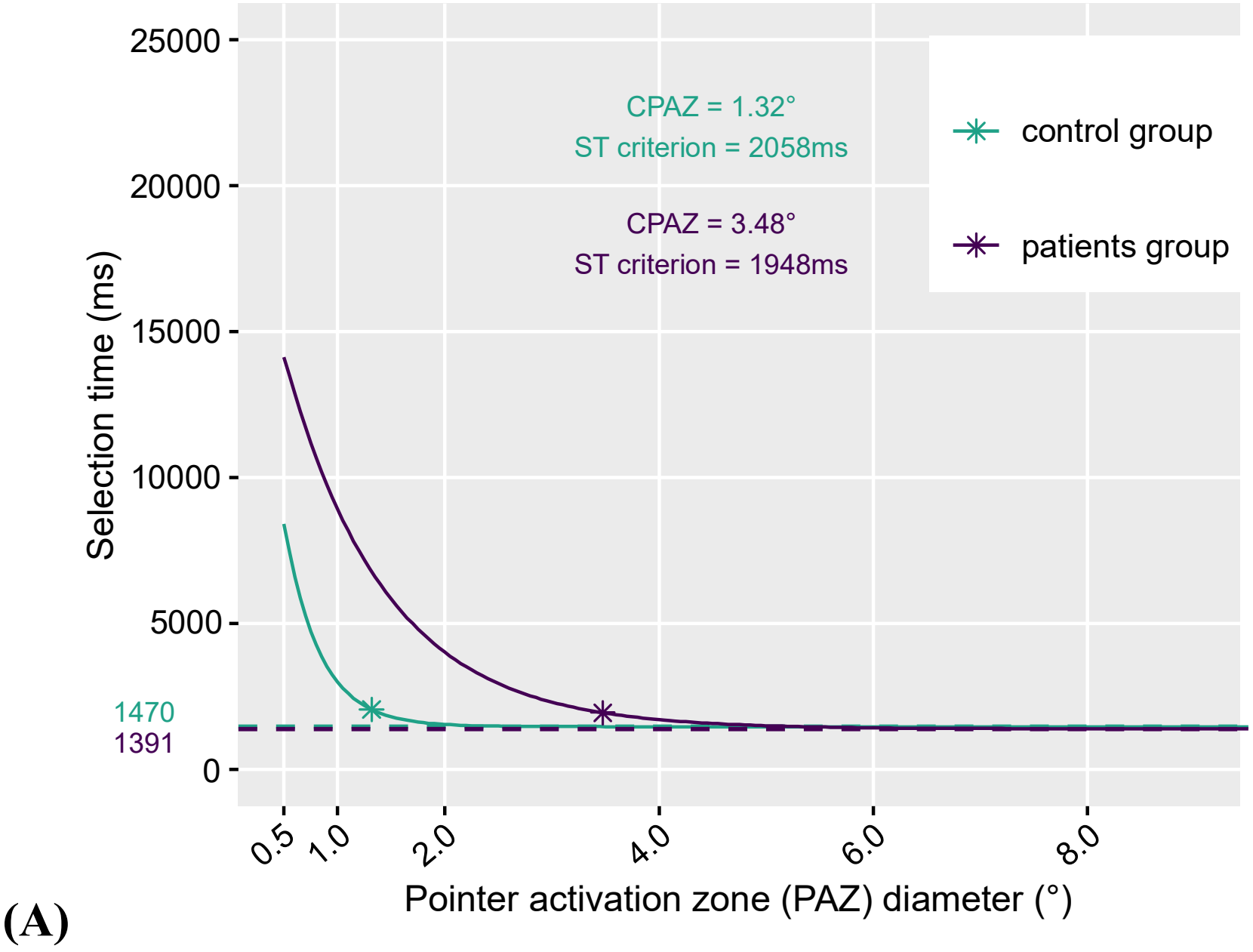

**(A)**

**(B)**

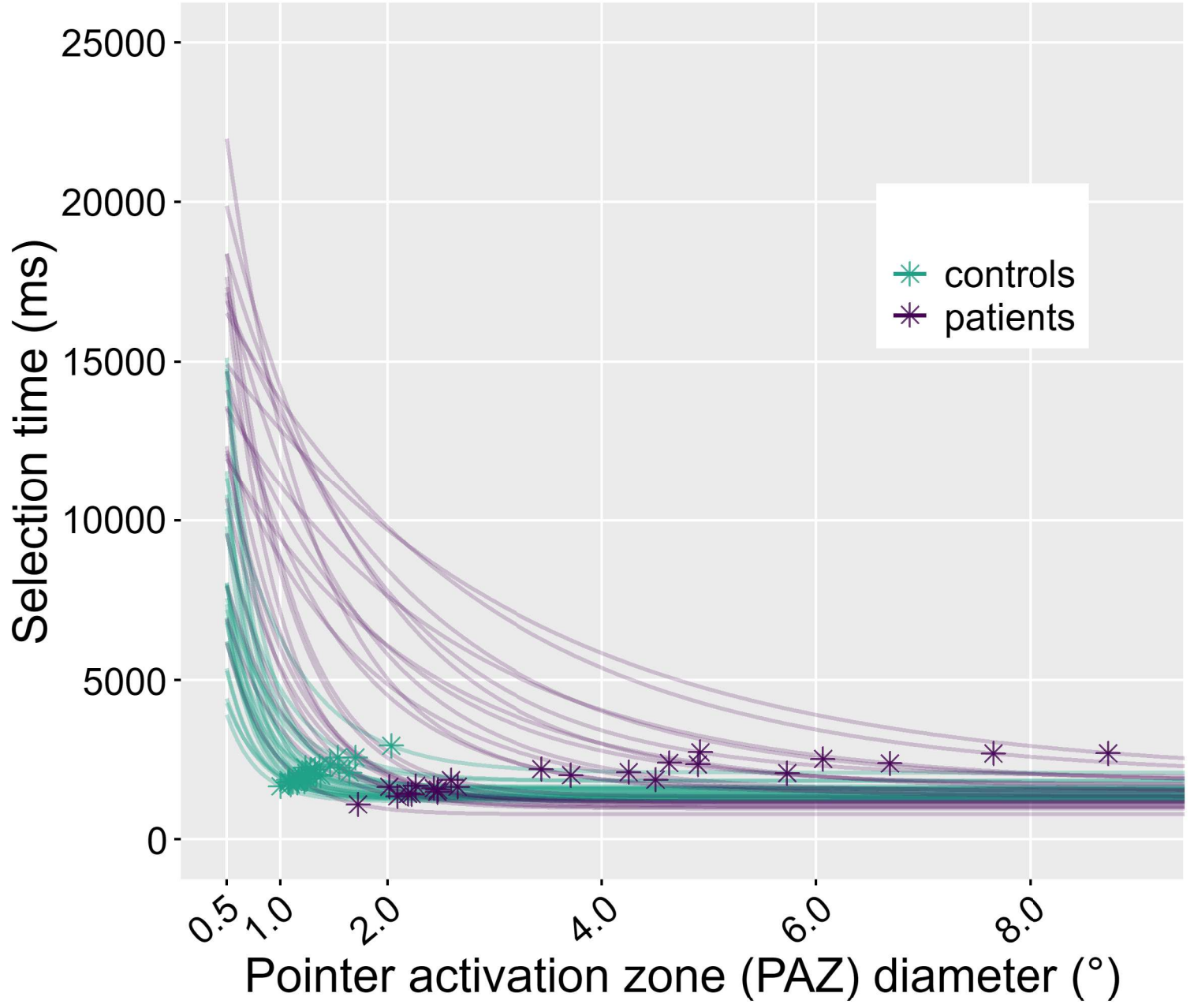

*Figure 9. CPAZ for the patients and control groups (A) and for each participant (B). CPAZ diameters are shown by dark purple stars in the patients group and light green stars in the control group. Solid lines represents the effect of PAZ on selection time from the NLMM (dark purple for the patients and light green for the controls). Dashed lines in (A) represent asymptote estimates at population-level.*

### 3.2. Reticle preference in the 3-reticle condition

In the 3-reticle condition, we measured the pourcentage of preference of each reticle position (ipsilateral, middle and contralateral). The fixed effects of the GLMM are plotted in Figure 10 and presented in Table 4. The estimated marginal means of the model are presented in Table 5. The effect of reticle position is significant ($p < .0001$). Pairwise comparisons show that patients and controls use significantly more the ipsilateral reticle (IPSI: patient = 71.9 % ; controls = 80.5 %) than the middle reticle (MID: patient = 21.4 % ; controls = 17.2 %) and the contralateral reticle (CONTRA: patient = 6.7 % ; controls = 2.4 %), all $p < .0001$. They also use more the middle reticle than the contralateral reticle ($p < .0001$). The interaction between group and reticle position is significant ($p < .0001$). Pairwise comparisons show that controls have a higher preference for the ipsilateral reticle than patients ($p <.0001$). Conversely, patients use more the middle reticle ($p = .0062$) and the contralateral reticle ($p < .0001$) than controls. This difference in the reticle preference is illustrated in Supplementary figure S1 and Supplementary figure S2 where the percentage of preference of each reticle is presented separately for each participant of the patients and control groups. The majority of control participants used only the ipsilateral reticle while some patients used more the middle or the contralateral reticles rather than the ipsilateral reticle.

To sum-up, both groups showed a strong preference for the ipsilateral reticle rather than the contralateral or the middle reticule. The preference for the ipsilateral reticle was stronger for the controls than patients, the later using more the middle and contralateral reticles than the controls.

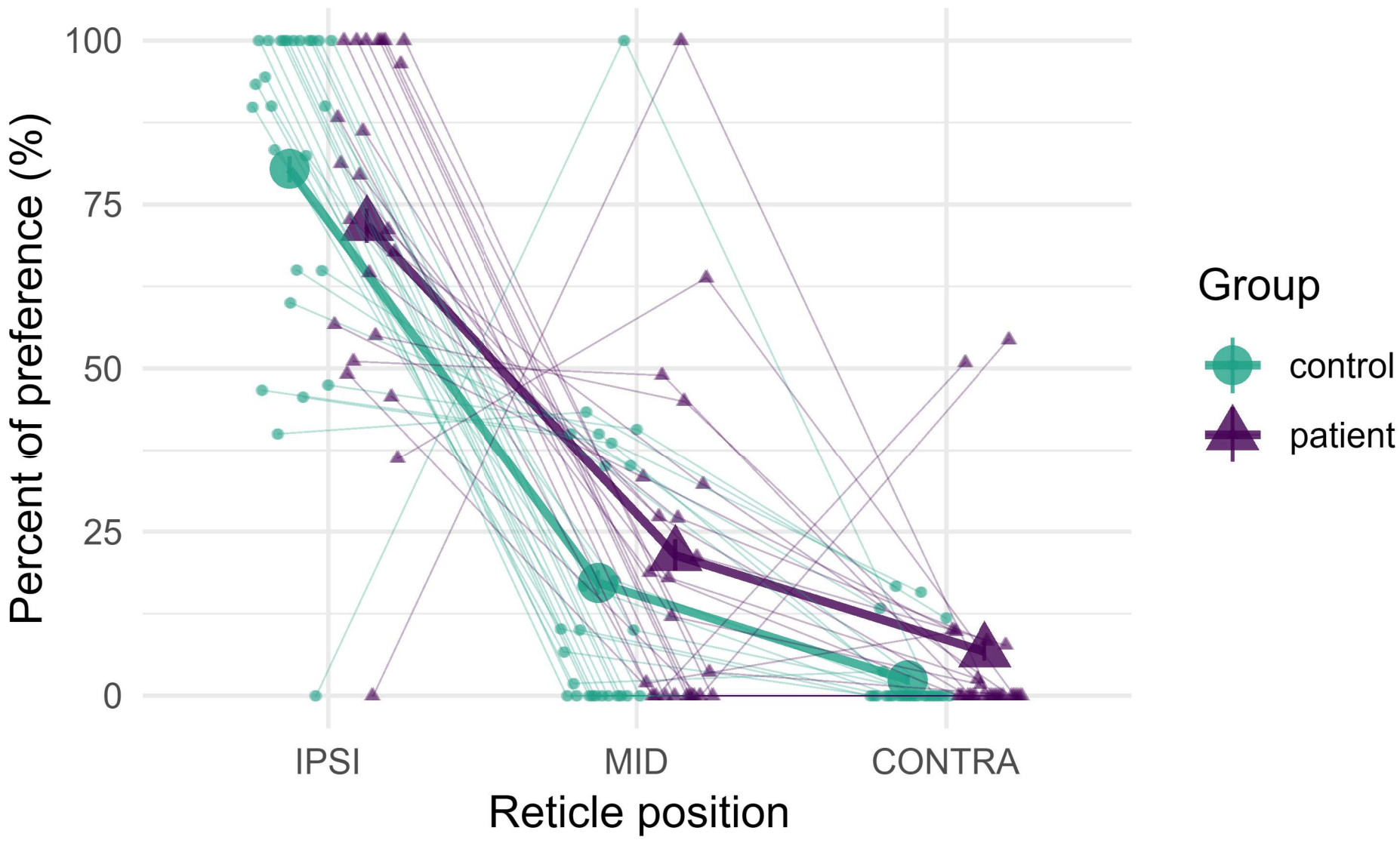

*Figure 10. Effect of the position of the reticle and group on the preference of the reticle from the GLMM. Large points and lines represent the effect of the position of the reticle at the population level for the patients group (dark purple triangle) and the control group (light green dot). Error bars show the confidence interval. Small dots and thin lines represent the across-trial average percentage of preference measured for each participant. IPSI: ipsilateral, MID: middle, CONTRA: contralateral.*

*Table 4. Fixed effects estimates from the generalized linear mixed-effects model. Estimates are given in log-odds. The reference level of the « position » and « group » independant variables are respectively the ipsilateral position (IPSI) and the control group. MID: middle position. CONTRA: contralateral position. SE : standard error. ***: p-value < .0001.*

| | Estimate | SE | z-value | p-value |
|---|---|---|---|---|
| Intercept (PositionIPSI) | 1.41442 | 0.06461 | 21.892 | *** |
| PositionMID | -2.98653 | 0.09372 | - 31.867 | *** |
| PositionCONTRA | -5.13609 | 0.18088 | - 28.395 | *** |
| GroupPatient | -0.47697 | 0.09250 | - 5.157 | *** |
| PositionMID × GroupPatient | 0.74904 | 0.13573 | 5.519 | *** |
| PositionCONTRA × GroupPatient | 1.56902 | 0.22645 | 6.929 | *** |

*Table 5. Estimated marginal means from the generalized linear mixed-effects model. Marginal means from the GLMM are given both in log-odds with their standard error (SE), and in % with their confidence interval (CI).*

| | Marginal means (log-odds) | SE (log-odds) | Marginal means (%) | CI (%) |
|---|---|---|---|---|
| Control – IPSI | 1.414 | 0.0646 | 80.45 | [78.38, 82.36] |
| Control – MID | -1.572 | 0.0679 | 17.19 | [15.38, 19.17] |
| Control – CONTRA | -3.722 | 0.1689 | 2.36 | [1.71, 3.26] |
| Patient – IPSI | 0.937 | 0.0662 | 71.86 | [69.16, 74.41] |
| Patient – MID | -1.3 | 0.0725 | 21.42 | [19.12, 23.9] |
| Patient – CONTRA | -2.63 | 0.1188 | 6.73 | [5.4, 8.34] |

## 4. Discussion

The general context of the present work is that low vision individuals face serious issues when interacting with complex digital interfaces, for instance to select items in a menu[39,40]. In this context, our general goal is to improve visually-guided pointing when low vision individuals with CFL, or more generally with visual impairments, are engaged in complex human-machine interactions. More precisely, the main goal of this study was to psychophysically investigate with these persons the efficiency of some techniques already used to improve pointing in normally-sighted persons. Performance was assessed by measuring the time to select a target. Individuals with CFL and normally-sighted observers were tested with a head-contingent pointing task in a virtual environment (the visible component of this pointing technique, or cursor, was a 6° diameter reticle – see Figure 1).

An important feature of our work is that the pointing technique used to point at targets is derived from the "area cursor" concept initially developed with normally-sighted persons[46]. We refer to this construct as the Pointer Activation Zone – PAZ - see Figure 1) to make its role more explicit, and most importantly to show that the visual appearance of our cursor (here a reticle) is independent from the PAZ characteristics. In short, increasing the size of the PAZ decreases the difficulty of the pointing task while keeping the visual appearance of the reticle constant. Basically, increasing the PAZ size amounts to decreasing the degree to which the cursor is required to be aligned with the center of the target object, a relationship consistent with the famous Fitts' law[45]. Benefits of area cursors have been shown with normally-sighted individuals[44,46] and people with motor impairments[43]. To our knowledge, the present study is the first to extend this pointing facilitation effect to individuals with CFL. Note also that the use of a reticle having a very large angular size (here 6°) is an important feature of our experimental paradigm that might also be beneficially incorporated in future VR pointing interfaces. The key advantage of this large reticle is that it does not provide visual masking effects in the vicinity of the target while providing an efficient pointing interface as shown by our results.

Our first main result is that selection time decreases with increasing PAZ diameter both for patients and controls. In other terms, pointing performance increases with decreasing task difficulty. A non-linear, asymptotic, relationship between task difficulty and selection times is observed. Patients need more time than controls to select the target when PAZ diameter is the smallest (R0 estimate). The best performances (asymptote estimate) of patients and controls, namely around 1400 ms (the 1500 ms dwell-time not being included), are not significantly different. However the rate of decrease towards the asymptote (lrc estimate) is smaller for patients than controls. This means that patients' best performance can be equated to controls' best performance only with the largest PAZ diameter values of the study. The drawback of these large PAZ values is that, despite their quicker response time, they induce relatively poor pointing resolution. A compromise between pointing resolution and selection time can be found if PAZ diameter is chosen to be large enough to provide reasonably low selection time (near the asymptotic level), but not too large to avoid poor pointing resolution. In that perspective, we defined the concept of Critical Pointer Activation Zone (CPAZ), that is a PAZ diameter threshold value above which selection time starts remaining below a criterion value slightly above the asymptote estimate. Choosing an individual CPAZ value might be based on results obtained at the individual-level or at the population-level if time is missing for individual measures - see Figure

9. Apart from optimizing pointing performance, extracting individual CPAZ values has the advantage of characterizing each individual's performance with a single value. This is similar for instance to the use of the Critical Print Size used in many reading studies especially to characterize and compare reading performance of low vision persons[64].

In our work, inspection of Figure 11 suggests that CPAZ values increase as a function of visual acuity only for the patients group. This is confirmed by a linear model showing that the interaction between visual acuity and group is significant ($p = .025$) although the effect of visual acuity is not significant. This clear relationship between CPAZ and acuity is important as it suggests that the parametric pointing task developed in our work relies in large part on low-level visual processes (as compared notably to motor processes). This low-level signature of our pointing task is also suggesting that practicing pointing tasks might be a new practical and effective tool for the rehabilitation of patients with CFL (see below our discussion on the use of rehabilitation games).

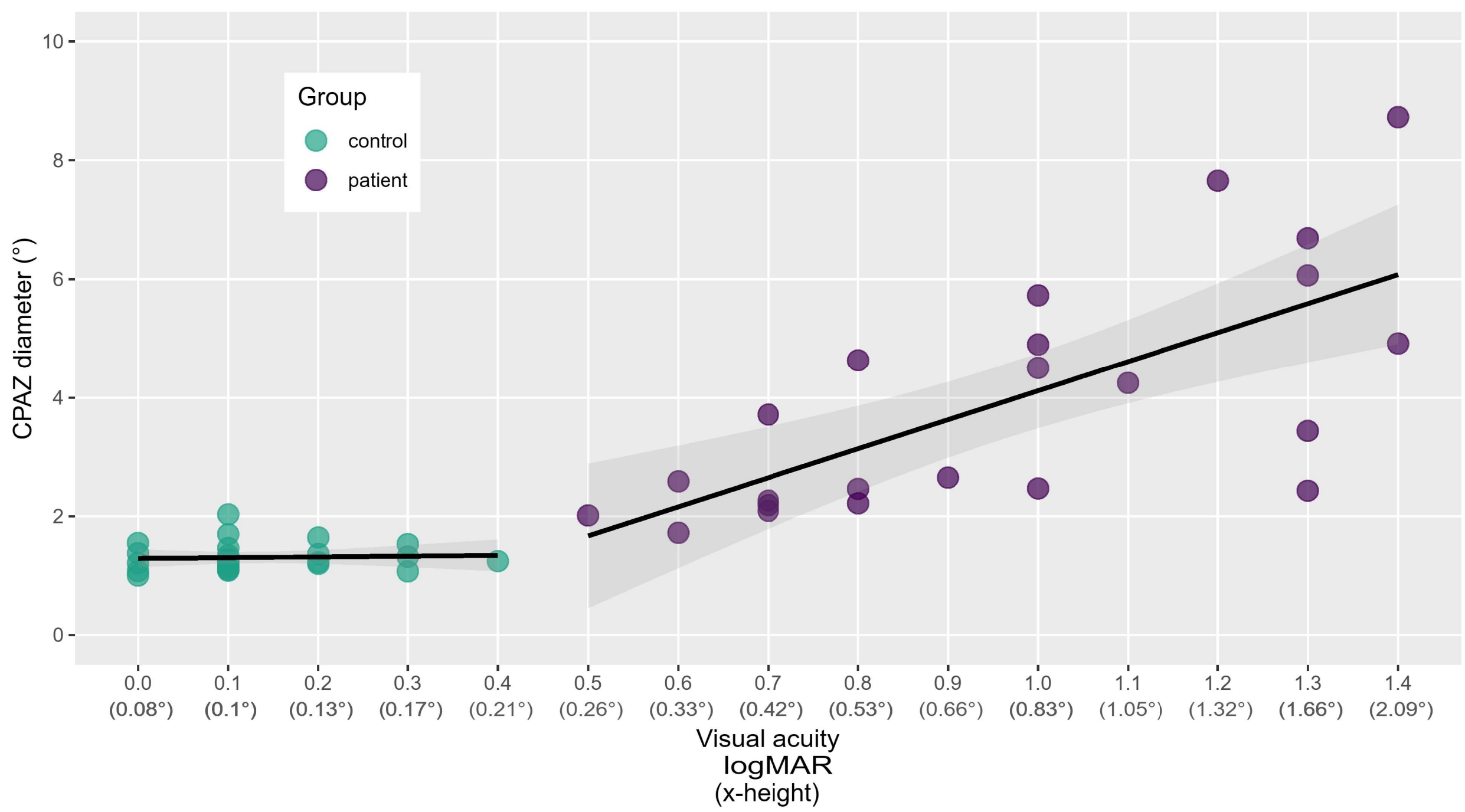

*Figure 11. Critical Pointer Activation Zone (CPAZ) diameter as a function of visual acuity of the viewing eye (worst and best eye respectively for patients and control groups). CPAZ diameter values are shown for each patient (dark purple dots) and for each control participant (light green dots). The black line represents the smoothing curve across controls and patients using a linear regression. Visual acuity is given in logMAR along with the corresponding letter x-height in degrees.*

Our second main result concerns the following question : is the distance between the target and the reticle an important aspect of pointing performance? This issue was investigated in independent experimental blocks where participants were tested with three simultaneously-displayed reticles (hence three pointer activation zones) compared to a single reticle (Figure 2A). In the 3-reticle condition, participants were free to choose at any time the reticle they wanted to use for pointing. Our hypothesis was that participants would prefer using the closest cursor to the target. In keeping with this prediction, we find that patients with CFL and normally-sighted individuals have a strong preference for using the reticle closer to the target (ie, the reticle ipsilateral to the target) rather than one of the two more distant reticles (contralateral or middle position). The closer reticle (left or right) is used in 71.9 % of the trials by patients, and 80.45% by controls. The middle reticle is less often used (patients: 21.4% ; controls: 17.2%), but still more often than the furthest reticle (patients: 6.7%;

controls: 2.4%). It is plausible that participants preferred using the ipsilateral reticle because it minimized the required amplitude of head movements. Despite this preference, analyses of selection times showed that presenting three cursors instead of a single one (as used in standard pointing tasks) did not lead to an improvement of performance. Parameters of the psychometric curves for selection times were not significantly different between the 1- and 3-reticle conditions, neither for patients nor controls. This pattern of results might give perspectives in the context of improving pointing techniques in complex human-machine interactive digital interfaces. For example, when a menu is displayed on the left side of the screen, it would be relevant to display the cursor also on the left with a smaller eccentricity. This principle, based on the optimization of target-cursor proximity, could be easily implemented in digital interfaces and might also be dynamic: for instance, the sudden appearance of a leftward menu could concomitantly induce the appearance of a leftward cursor. Actually, cursors that dynamically change across time have already shown benefits[44]. Our results suggest that this technique might be helpful to reduce the fatigue or discomfort induced by pointing movements (here head movements), which would be an interesting asset for patients even in the absence of any selection time improvement.

Overall, these results are encouraging as they provide clear evidence that persons with CFL are able to efficiently perform pointing tasks in virtual reality environments. The limits of pointing performance can be parametrically investigated with the psychophysical approach developed in the current study. This psychophysical experimental paradigm might be used in the future to investigate the effect of factors that might potentially improve pointing performance for low vision persons. With this aim in mind, a large set of data on these factors with normally-sighted individuals is already available and offers many perspectives : various factors have been studied, for example target size[45], target distance from the pointer[45], effective target size manipulated with the « area cursor »[46], the body part used to point at objects[69], pointing validation techniques (e.g. point-and-dwell or point-and-click[70]), dwell time[48], to name a few. As a reminder, in the current study, a head-contingent reticle cursor with a variable pointer activation zone was provided to point at the target, a visual feedback (visual flickering) was used to indicate if the target was correctly pointed at (« pre-selection » period), and pointing was validated after a dwell time of 1500 ms (actual selection). Among the numerous factors already investigated with normally-sighted individuals, we emphasize some of them as they might be a logical continuation of the present investigation. For instance, point-and-dwell methods have the potential drawback to induce unintentional selection (similar to the Midas Touch Problem – see [71]). One possibility to solve this problem is to replace the point-and-dwell method by point-and-click methods where the pre-selected target is eventually validated by pressing a button (e.g. the trigger of a handcontroller). In the same vein, the nature of the feedback used to indicate if the target is correctly pointed at during the pre-selection period is an important factor to investigate further. While the visual flickering feedback used in our study was intended to optimally activate the peripheral retina of individuals with CFL[72], it would be interesting to investigate if other kinds of feedback might be more efficient.

Improving the accessibility of pointing tools for persons with low vision is all the more important that pointing is a key component of some innovative visual aids whose principle is to smartly augment some specific Regions Of Interest (ROIs) (e.g. [73]). This "Select and Augment Segmented Items" (SASI) strategy has been initially shown to improve performance when reading with a simulated macular scotoma[53]. With low vision patients, SASI algorithms have also been shown to improve face recognition[54,55] or watching videos[74]. We also believe that visually-guided pointing tasks, whether associated with SASI algorithms or not, could be a key component of low vision rehabilitation protocols if they were rendered more accessible. This is especially true for rehabilitation protocols that aim at combining tasks involving low-level, oculomotor and attentional processes, all of which are significantly involved in pointing tasks for persons with low vision[75–78]. This could be achieved by creating serious games (e.g. [79,80]) relying on an extensive practice of visually-guided pointing tasks where the difficulty of these tasks would be minor for low vision patients. More precisely, the game difficulty could be adapted to each patient's threshold by varying some key parameters such as the PAZ diameter. To this aim, we have been implementing VR games where players must constantly produce a multitude of easy-to-make pointing gestures. These games were developed with the open-source PTVR toolbox[52] (https://ptvr.inria.fr/). They are freely available to allow low vision practitioners and vision scientists to use them without restriction. The specificity of these VR games is that they contain : a/ stimuli of very large angular sizes (impossible to obtain with standard 2D displays), and b/ tools to remove the difficulty of pointing (such as the PAZ investigated here or SASI-similar tools). The easiness of these games for individuals with CFL also relies on the fact that pointing tasks only require simple and minimal instructions. A typical example of the games we have developed is the « balloon burst » game that we initially designed for children with low vision. A typical game for adults is the « 4 images 1 word » game where letters must be constantly dragged and dropped (thus implying typical pointing gestures) to form words. We are currently evaluating in pilot studies how these games are perceived by individuals with CFL.

**Conclusion**

Our work uses a psychophysical analysis of pointing performance in virtual reality with the hope of helping design more accessible pointing techniques in digital interfaces for low vision individuals with CFL. Our study shows that point-and-dwell selection techniques using a large visible reticle associated with a Pointer Activation Zone – PAZ (based on the initial "area cursor" concept-[46]) are accessible for low vision users. Furthermore, our results indicate that accessibility of pointing techniques can be improved by customizing the PAZ with an optimal size. This optimal threshold size might be extracted by measuring each individual's Critical PAZ (CPAZ) value, or by using published CPAZ values calculated at the population level. This could be fruitfully applied for instance to develop interfaces allowing users to select an item on a menu containing several items.

Given that factors facilitating pointing for low vision individuals with CFL are still understudied, further investigations must be conducted to better understand the limiting factors of pointing for these individuals. The present study offers a psychophysical paradigm that could be used to conduct these studies in a controlled and homogeneous framework.

More generally, our study highlights the potential of VR (thanks to large angular values and rich interactivity) to develop more inclusive pointing techniques for low vision individuals. VR is also the perfect medium to develop easy-to-understand and easy-to-play pointing-based games specifically designed for persons with low vision. In the future, all these techniques might be incorporated into Augmented Vision systems.

**Acknowledgments**

The authors thank all the participants who were volunteered to take part in this study. This work was supported by grants from the Carnot Cognition Institute, the NeuroMarseille Institute and the ANR (20-CE19-0018).

## 6. Supplementary figures

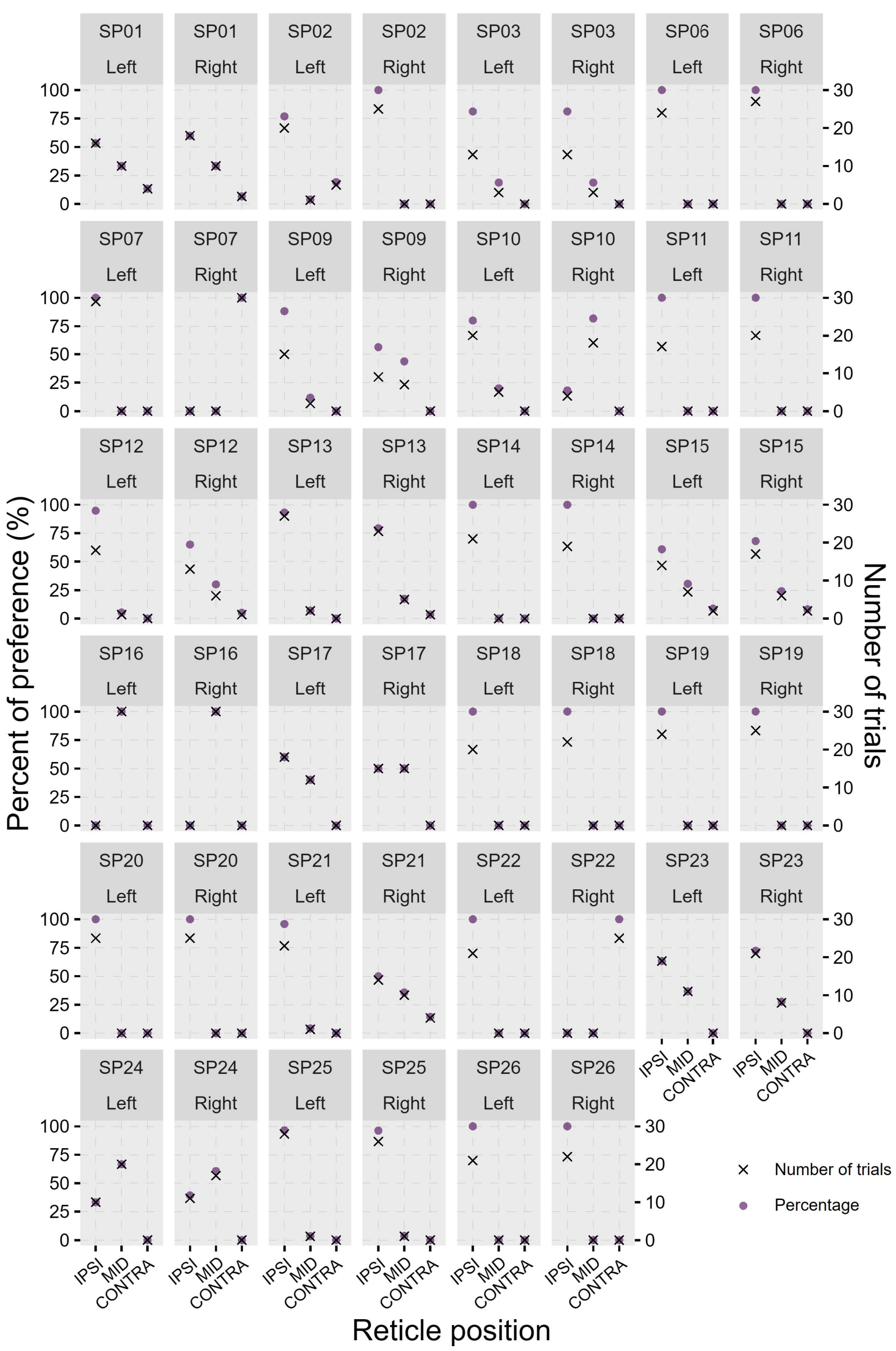

*Supplementary figure S1. Repartition of the prefered reticles in the 3-reticle condition for each patient. Each facet represents one participant when the target was on the left or right side. The repartition of preference between the ipsilateral (IPSI), contralateral (CONTRA) and middle (MID) reticle positions is given in percent (purple dots, left axis) and number of trials (black crosses, right axis).*

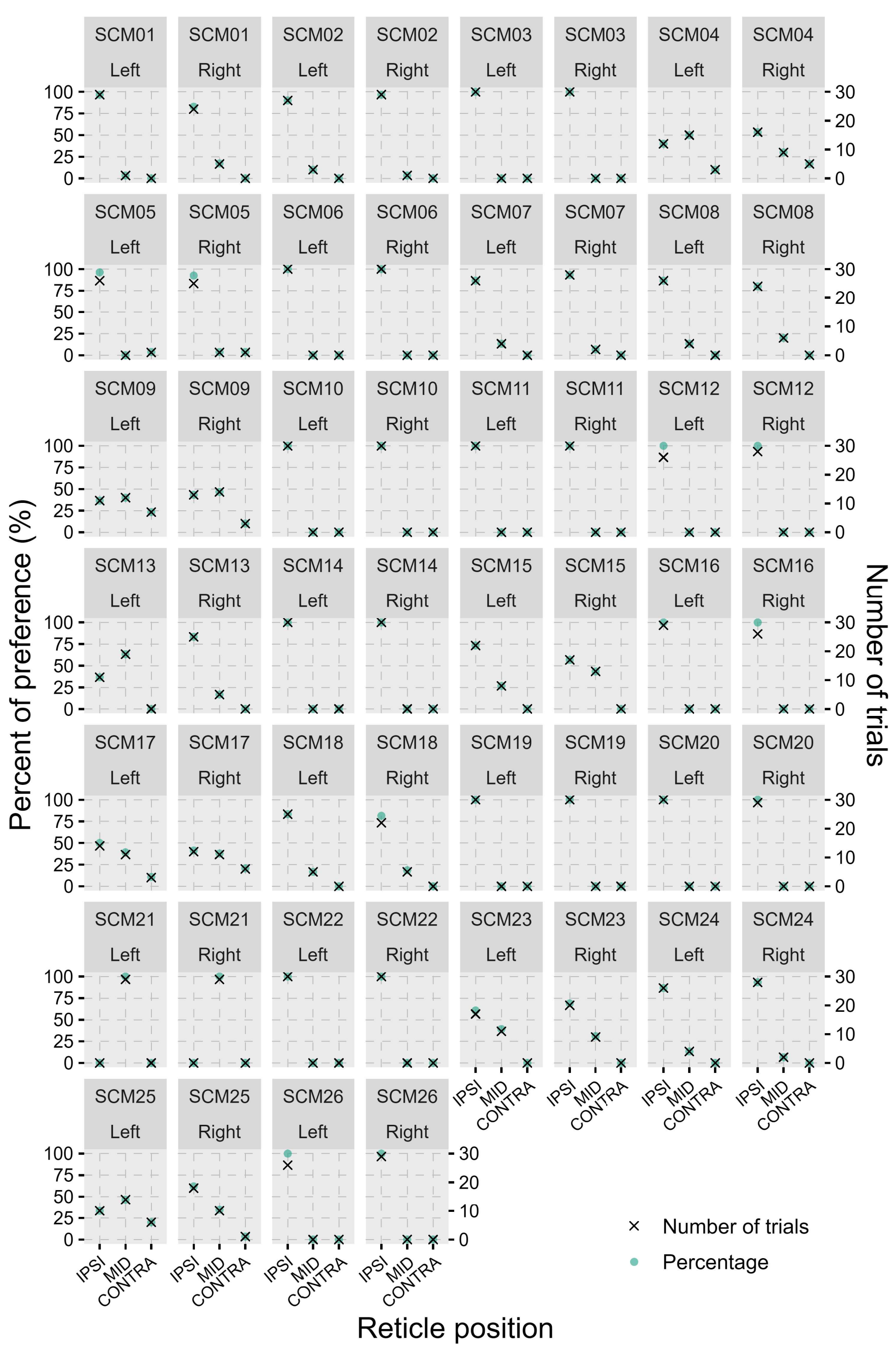

*Supplementary figure S2. Repartition of the prefered reticles in the 3-reticle condition for each control. Each facet represents one participant when the target was on the left or right side. The repartition of preference between the ipsilateral (IPSI), contralateral (CONTRA) and middle (MID) reticle positions is given in percent (green dots, left axis) and number of trials (black crosses, right axis).*